\documentclass[numberedappendix]{emulateapj}
\usepackage{graphics,graphicx}
\usepackage{helvet}
\usepackage[utf8]{inputenc}
\usepackage[T1]{fontenc}
\usepackage{longtable}
\usepackage{soul}
\usepackage{hyperref}
\usepackage{amsmath, amssymb, gensymb}
\usepackage{ulem}
\usepackage{natbib}
\usepackage{microtype}
\usepackage{multirow}
\newcommand{\mjybm}{mJy~beam$^{-1}$}
\newcommand{\jybm}{Jy~beam$^{-1}$}

\newcommand{\etal}{{\it et al.\ }}

\begin{document}

\slugcomment{{\it Accepted for publication in AJ: February 15, 2013.}}

\title{Interferometric Upper Limits on Millimeter Polarization of the Disks around DG Tau, GM Aur, and MWC 480}
\shorttitle{Disk Polarization Limits}

\author{A. Meredith Hughes\altaffilmark{1,2}, 
Charles L. H. Hull\altaffilmark{1},
David J. Wilner\altaffilmark{3},
Richard L. Plambeck\altaffilmark{1}
}

\shortauthors{Hughes \etal}
\email{amhughes@wesleyan.edu, chat@astro.berkeley.edu}

\altaffiltext{1}{Astronomy Department \& Radio Astronomy Laboratory, University of California, Berkeley, CA 94720-3411, USA}
\altaffiltext{2}{Van Vleck Observatory, Astronomy Department, Wesleyan University, 96 Foss Hill Drive, Middletown, CT 06459, USA}
\altaffiltext{3}{Harvard-Smithsonian Center for Astrophysics, 60 Garden St., Cambridge, MA 02138, USA}


\begin{abstract}
Millimeter-wavelength polarization measurements offer a promising method for probing the geometry of magnetic fields in circumstellar disks.  Single dish observations and theoretical work have hinted that magnetic field geometries might be predominantly toroidal, and that disks should exhibit millimeter polarization fractions of 2--3\%.  While subsequent work has not confirmed these high polarization fractions, either the wavelength of observation or the target sources differed from the original observations.  Here we present new polarimetric observations of three nearby circumstellar disks at 2$\arcsec$ resolution with the Submillimeter Array (SMA) and the Combined Array for Research in Millimeter Astronomy (CARMA).  We reobserve GM~Aur and DG~Tau, the systems in which millimeter polarization detections have been claimed.  Despite higher resolution and sensitivity at wavelengths similar to the previous observations, the new observations do not show significant polarization.  We also add observations of a new HAeBe system, MWC~480.  These observations demonstrate that a very low ($\lesssim$\,0.5\%) polarization fraction is probably common at large ($\gtrsim$\,100\,AU) scales in bright circumstellar disks.  We suggest that high-resolution observations may be worthwhile to probe magnetic field structure on linear distances smaller than the disk scale height, as well as in regions closer to the star that may have larger MRI-induced magnetic field strengths.
\end{abstract}

\keywords{magnetic~fields -- polarization -- protoplanetary disks -- stars: formation -- stars: individual(DG~Tau,~GM~Aur,~MWC~480)}

\section{INTRODUCTION}
\label{sec:intro}

Magnetic fields are frequently invoked in theories of star and planet formation; \citet{kon11} provide a thorough review of the central role of fields in the formation and evolution of circumstellar disks.  While some observations find hourglass-shaped magnetic field geometries in very young star forming regions, which indicates that magnetic fields play a key role in the early formation of circumstellar disks \citep[e.g.,][]{gir06}, there are few observational constraints on magnetic field configurations at later stages.  Observations of polarized, millimeter-wavelength continuum emission are ideally suited for determining magnetic field morphology independent of circumstellar disk geometry \citep{ait02}.  Polarized thermal emission is generated by elongated dust grains aligned with the magnetic field.  Under most conditions, grains are expected to rotate with their long axes perpendicular to the direction of the magnetic field, so that they generate polarized emission perpendicular to the magnetic field direction (see review in \citet{laz07}, as well as \citet{hoa09}).  A set of encouraging observations with the JCMT resulted in tentative detections of polarized 850\,$\mu$m emission from two T Tauri stars, GM Aur and DG Tau \citep{tam99}.  In both cases, the position angle of the polarization vector was consistent with the orientation of the minor axis of the disk, suggesting that the magnetic fields are toroidal.  Because the 15$\arcsec$ beam of the JCMT is far larger than the few-arcsecond sizes typical of circumstellar disks, only a single polarization vector could be detected.  This result was followed up with theoretical work by \citet{cho07}, who presented a theoretical justification for the observed globally averaged 2--3\% polarization fraction, and predicted that it should be common to disks around young stars.  The \citet{cho07} investigation resulted in a theoretical framework that can be used to model the position-dependent fractional polarization of emission from circumstellar disks.

Subsequent observations have not confirmed the expected 2--3\% polarization fraction.  \citet{kre09} revisited the DG Tau disk at a wavelength of 350\,$\mu$m with the Caltech Submillimeter Observatory (CSO) and did not detect polarized emission in the 9$\arcsec$ beam, placing a 2$\sigma$ upper limit on the integrated fractional polarization of 1.7\%.  Roughly concurrent interferometric  observations of the disks around HD 163296 and TW Hya, selected for their large millimeter fluxes so as to provide the best sensitivity to fractional polarization, also did not detect any polarized emission at a wavelength of 880\,$\mu$m \citep{hug09b}.  Those observations ruled out the fiducial  \citet{cho07} model at the 10$\sigma$ level, with an estimated integrated  upper limit on the polarization fraction of $\lesssim$\,0.5\% (since the interferometric observations spatially resolve the disk, comparisons with integrated polarization fractions from single-dish telescopes are necessarily model-dependent).  \citet{hug09b} also explored the theoretical parameter space to determine which factors are most likely to cause the suppression of polarized emission in the observed systems, and determined that some combination of rounding of large grains, reduced efficiency of alignment of large grains with the magnetic field, and/or some degree of magnetic field tangling (perhaps due to turbulence) could plausibly explain the low polarization fraction.

Some ambiguity remains as to whether the \citet{kre09} and \citet{hug09b} results apply to the \citet{tam99} results.  In the former case \citep{kre09}, the wavelength of the DG Tau observation was different enough that the non-detection of polarization could result from wavelength-dependent processes like scattering.  In the latter case \citep{hug09b}, both the lack of overlap in targets and the small number of sources introduce uncertainty regarding how broadly applicable the low polarization fraction might be.  We have re-observed the two sources that have polarization detections from \citet{tam99}, plus another bright HAeBe system, MWC 480.  These new observations more than double the sample of circumstellar disks with interferometric, millimeter-wavelength polarization observations, and help to clarify the results of previous observations by revisiting sources with single-dish detections at the same wavelength, but with higher sensitivity and spatial resolution.

Here we present new 880\,$\mu$m polarimetric observations of the circumstellar disks around GM Aur and MWC 480 with the Submillimeter Array (SMA), and 1.3\,mm observations of the disk around DG Tau with the Combined Array for Research in Millimeter Astronomy (CARMA).  In Section~\ref{sec:results} we compare our observations with the \citet{cho07} predictions for the spatially resolved distribution of polarized flux in T Tauri disks, as well as with the overall polarization fraction detected by the JCMT for the GM Aur and DG Tau systems.  In Section~\ref{sec:discussion} we discuss the implications of our results for our understanding of magnetic fields in T Tauri stars, and we summarize our work in Section~\ref{sec:summary}.

\section{OBSERVATIONS AND DATA REDUCTION}
\label{sec:observations}

\subsection{SMA observations}

We conducted 880\,$\mu$m observations of GM~Aur with the SMA polarimeter \citep{mar08} on the night of 2009 November 6.  Our method was very similar to that described in \citet{hug09b}.  The array was in the compact-north configuration, with baseline lengths between 16 and 123\,m, and the longest baselines along the north-south direction.  The weather was good, with 225\,GHz opacity of 0.07 and a stable atmospheric phase.  MWC~480 was observed on the night of 2009 December 11 in similarly good weather, with stable phase and a 225\,GHz opacity of 0.06.  The array was in the compact configuration (16--77\,m baselines) for the MWC~480 observations.

The data were collected with the receivers tuned to a local oscillator (LO) frequency of 341\,GHz (880\,$\mu$m wavelength).  These observations differ from those described in \citet{hug09b} in that a new high-bandwidth observing mode was commissioned on the SMA in the intervening period, which effectively doubled the bandwidth of each sideband from 2\,GHz to 4\,GHz.  We use a uniform correlator configuration that divides each 104\,MHz-wide correlator chunk into 128 channels to achieve maximum continuum sensitivity at the highest possible uniform spectral resolution (0.7\,km\,s$^{-1}$).

For both GM~Aur and MWC~480, the quasar 3c111 served as the atmospheric and instrumental gain calibrator, and J0510+180 was included in the observing loop to test the quality of the phase transfer.  3c111, J0510+180, Callisto, 3c273, and 3c279 were used as passband calibrators.  Callisto was observed near the beginning of each track to determine the absolute flux scale, and Titan was also included at the end of the MWC~480 track; we derive fluxes of 1.09 and 1.52\,Jy for 3c111 on the nights of November 6 and December 11, respectively.  The instrumental polarization calibration was carried out as described in \citet{hug09b}, by observing 3c273 and 3c279 in the hours before and after transit.  We derive consistent instrumental leakage solutions for both sources, and adopt the solutions from the brighter 3c273.  The passband, gain, and instrumental polarization calibrations were carried out independently for each 2\,GHz segment of each sideband.  A summary of the observational parameters is given in Table~\ref{tab:obs}.  

The data were edited and calibrated with the MIR software package\footnote{See \url{http://cfa-www.harvard.edu/~cqi/mircook.html}.}, and the standard tasks of inversion and beam deconvolution were carried out using the \texttt{MIRIAD} software package.

\subsection{CARMA observations}

We conducted $\lambda$1.3\,mm polarimetric observations of DG~Tau at CARMA in the C-array (30--350~m baselines) on 2012~March~3, and in the D-array (11--50~m baselines) on 2012~October~13. The data were taken in average weather, with 225 GHz opacities of 0.28 and 0.26, and median atmospheric phases of 90 and 165~$\mu$m for the C and D-array data, respectively.  These observations were taken as part of the TADPOL survey, a CARMA key project described in \citet{Hull2012}.  

The data were collected with the receivers tuned to an LO frequency of 223.821\,GHz (1.3\,mm wavelength). We use a correlator configuration with three 500 MHz wide bands to measure the dust continuum, and one 31 MHz wide band to map spectral-line emission.  In this paper we focus only on the dust continuum polarization from the CARMA data.
 
The quasars 3c111 and J0510+180 served as the atmospheric and instrumental gain calibrators.  3c84, 3c111, and J0510+180 were used as passband calibrators.  Uranus was observed near the beginning of each track to determine the absolute flux scale.  Periods of unstable gains during both sets of observations led us to derive slightly lower than expected flux measurements for 3c111 and J0510+180, and thus low values for both the rms noise and the integrated flux of DG~Tau.  We instead adopt a flux of 317\,mJy for DG~Tau, based on a greater number of high-quality CARMA observations with overlapping baseline lengths presented in \citet{ise09}.  Using this value, we derive fluxes for J0510+180 of 2.14\,Jy on 3~March and 2.18\,Jy on 13~October; these flux values are consistent with the periods of greatest gain stability, as well as with archival flux measurements from the SMA obtained within weeks of the CARMA observations. The adopted fluxes are consistent with the derived fluxes within the typical 20\% systematic flux uncertainty generally expected for CARMA observations at $\lambda$1.3\,mm.  The adopted flux values also represent a conservative choice for the flux scale: the lower values derived from periods of unstable gains would not affect the measured fractional polarization, but would artificially decrease the measured rms noise in the maps.  Note that the peak Stokes I flux density is lower in C array than D array because the source is slightly resolved.

The leakage corrections are made in a similar manner to the SMA, by observing a strong source (either 3c111 or J0510+180, in these observations) over a range of parallactic angles.  Separate leakage solutions are made for the upper and lower sidebands.  Since CARMA has a simultaneous, dual-polarization system, an ``XYphase'' calibration is required to correct for phase differences between the left- and right-circular receivers.  This is done by observing an artificially polarized noise sources of known position angle.  

All data editing and calibration, as well as the standard tasks of inversion and beam deconvolution, were carried out using the \texttt{MIRIAD} software package.

\begin{table*}[hbt!]
\caption{Observational Parameters}
\footnotesize
\begin{center}
\begin{tabular}{lccccc}
\hline
 & GM~Aur & MWC~480 & \multicolumn{3}{c}{DG~Tau} \\
 & Com-N & Compact & C & D & C+D \\
\hline
\multicolumn{6}{c}{Continuum} \\
\hline
LO Frequency (GHz) & 341 & 341 & 224 & 224 & 224 \\
Beam Size (FWHM) & 2\farcs0$\times$1\farcs3 & 2\farcs6$\times$2\farcs1 & 0\farcs9$\times$0\farcs7 & 2\farcs3$\times$2\farcs2 & 1\farcs2$\times$0\farcs9 \\
~~~P.A. & 56$^\circ$ & 12$^\circ$ & -69$^\circ$ & 11$^\circ$ & -75$^\circ$ \\
RMS Noise (mJy\,beam$^{-1}$) & & & & & \\ 
~~~Stokes $I$ & 12 & 13 & 1.2 & 2.0 & 6.5 \\
~~~Stokes $Q$ \& $U$ & 2.8 & 3.1 & 0.71 & 0.84 & 0.66 \\
Peak Flux Density (mJy\,beam$^{-1}$) & & & & & \\ 
~~~Stokes $I$ & 360 & 880 & 220 & 290 & 250 \\
~~~Stokes $Q$ \& $U$ (3$\sigma$ upper limit) & $\le$\,8 & $\le$\,9 & $\le$\,2 & $\le$\,3 & $\le$\,2 \\
Integrated Flux (Stokes $I$; Jy) & 0.53 & 0.95 & 0.317 & 0.317 & 0.317 \\
\hline
\multicolumn{6}{c}{CO(3--2) line$^{a}$} \\
\hline
Beam Size (FWHM) & 2\farcs1$\times$1\farcs3 & 2\farcs5$\times$2\farcs2 & -- & -- & --\\
~~~P.A. & 75$^\circ$ & 35$^\circ$ & -- & -- & -- \\
RMS Noise (Jy\,beam$^{-1}$) & & & & & \\ 
~~~Stokes $I$ & 0.17 & 0.28 & -- & -- & -- \\
~~~Stokes $Q$ \& $U$ & 0.17 & 0.23 & -- & -- & --\\
Peak Flux Density (Jy\,beam$^{-1}$) & & & & & \\ 
~~~Stokes $I$ & 4.6 & 5.4 & -- & -- & --\\
~~~Stokes $Q$ \& $U$ (3$\sigma$ upper limit) & $\le$\,0.5 & $\le$\,0.7 & -- & -- & -- \\
Integrated Flux (Stokes $I$; Jy\,km\,s$^{-1}$) & 49 & 23 & -- & -- & -- \\
\hline
\end{tabular}
\end{center}
\tablerefs{{\it (a)}~Analyses of spectral line data were conducted for SMA only; CARMA has not yet been calibrated for making polarimetric line measurements.}
\label{tab:obs}
\end{table*}

\section{RESULTS \& ANALYSIS}
\label{sec:results}

We detect no polarized continuum emission from the disks around GM~Aur, MWC~480, or DG~Tau.  In the disks where we were able to perform polarimetric CO(3--2) observations (GM~Aur and MWC~480), we do not detect polarized molecular line emission.  The measured Stokes $I$ fluxes and Stokes $Q$ \& $U$ upper limits for the line and continuum observations are listed in Table~\ref{tab:obs}.  

As described in \citet{hug09b}, we interpret the polarimetric non-detections with the help of the simulations described in \citet{cho07}.  The \citet{cho07} code takes as inputs the stellar properties, viewing geometry, and parametric descriptions of the temperature and density structure of the disk, and then calculates the intensity and fractional polarization of thermal continuum emission as a function of position throughout the disk.  It assumes a purely toroidal magnetic field, with dust grains aligned via the radiative torque mechanism \citep[e.g.,][]{dol76,laz07}.  As described in \citet{cho07}, for disks with intermediate or high inclinations, the code typically predicts a global, integrated polarization fraction of 2--3\%, consistent with the \citet{tam99} observations; however, the polarization fraction varies with position across the disk such that the highest polarization fractions typically occur in the outermost regions of the disk where densities are lowest and grains are most readily aligned.  The \citet{cho07} code therefore provides a crucial interpretive mechanism that allows us to compare the polarization properties of the high-resolution interferometric observations with the low-resolution, integrated single-dish observations.  

We generate initial models of the disk temperature and surface density structure that allow us to reproduce the Stokes $I$ continuum data, based on stellar parameters drawn from the literature and on prior fits to millimeter-wavelength continuum observations by \citet{hug08,hug09}.  For DG~Tau, no recent power-law fit to high quality data exists.  There is a power-law fit to mid-1990s data from the Plateau de Bure Interferometer \citep{dut96} that has since been superseded by higher sensitivity observations; however, the fitting procedure used for more recent high-quality CARMA data \citep{ise09} cannot be reproduced in the simple parametric form needed for use with the code.  Instead, we compromise by constructing a simple power-law fit to the continuum emission in the following way: (1) we assume a disk temperature profile in which the grains are in radiative equilibrium with the central star, which is effectively identical to the temperature profile derived by \citet{dut96}; (2) we adopt an effective outer radius of 89\,AU based on the surface density fit of \citet{ise09}, which exhibits a very sharp drop in surface density at this radius; (3) we assume a surface density power law index $p=1$ based on the median value from the \citet{and09a} survey of disk structure in Ophiuchus (the data are only marginally resolved, and therefore of insufficient quality to distinguish between surface density power law indices); and (4) we then scale the normalization of the surface density to reproduce the observed integrated 1.3\,mm flux.  The power-law parameters adopted by this method are similar to both the \citet{dut96} and \citet{ise09} solutions.  We adopt an inclination for the DG Tau disk of 28.5$^\circ$, the average result from four data sets in \citet{ise10}.  The stellar and disk parameters used for modeling all three objects are listed with references in Table~\ref{tab:more_pol_model}.  Stellar parameters (temperature $T$, radius $R$, mass $M$) are indicated with an asterisk in the subscript.  $a$ denotes disk radii, inner ($a_{\mathrm{inner}}$) and outer ($a_0$).  The maximum grain size is $r_\mathrm{max,i}$, and the surface density at 100\,AU is denoted as $\Sigma_0$.  The adopted distance to the system $d$ is identical for all three sources, since they are all associated with the Taurus star forming region. 

\begin{figure*}
\epsscale{0.9}
\plotone{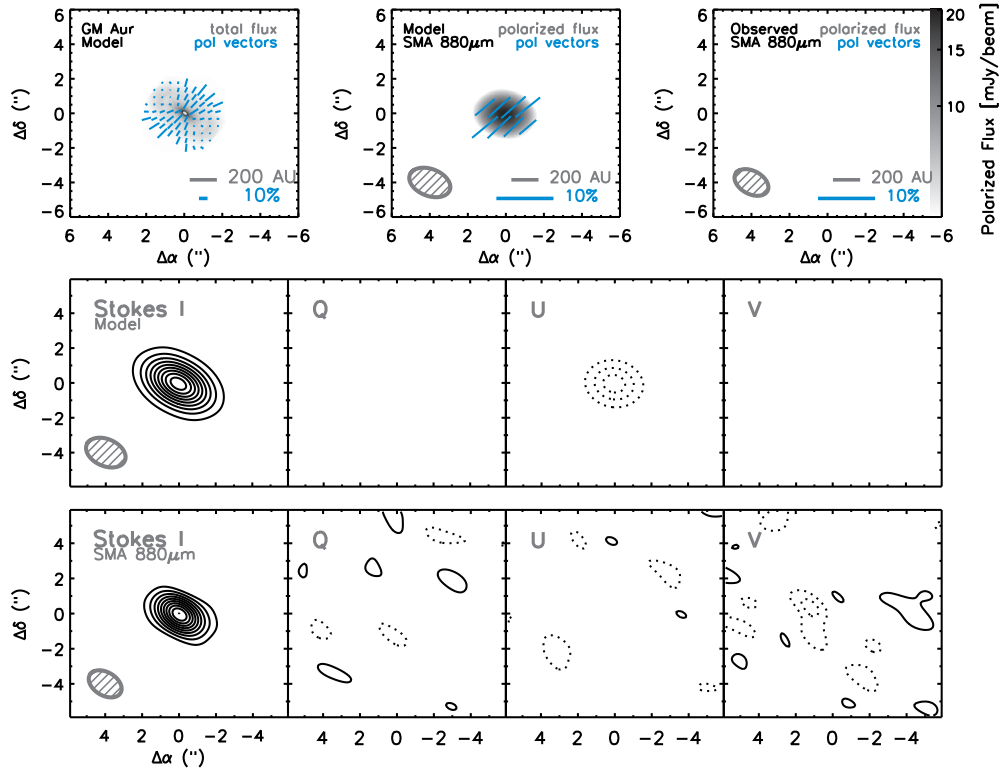}
\figcaption{ \footnotesize
Comparison of the \citet{cho07} model with the SMA 880\,$\mu$m observations of GM~Aur.  The top row shows the prediction for the model at full resolution (left), a simulated observation of the model with the SMA (center), and the SMA observations (right).  The grayscale shows either the total flux (left) or the polarized flux density (center, right), and the blue vectors indicate the percentage and direction of polarized flux at half-beam intervals.  The center and bottom rows compare the model prediction (center) with the observed SMA data (bottom) in each of the four Stokes parameters ($I$, $Q$, $U$, $V$, from left to right).  Contour levels are the same in both rows, either multiples of 10\% of the peak Stokes $I$ flux density (0.34 \jybm), or in increments of 2$\sigma$ for $Q$, $U$, and $V$, where $\sigma$ is the rms noise of 2.8 \mjybm.  The size and orientation of the synthesized beam is indicated in the lower left of each panel.
\label{fig:gmaur_pol}}
\end{figure*}

\begin{figure*}[t]
\centering
\epsscale{0.9}
\plotone{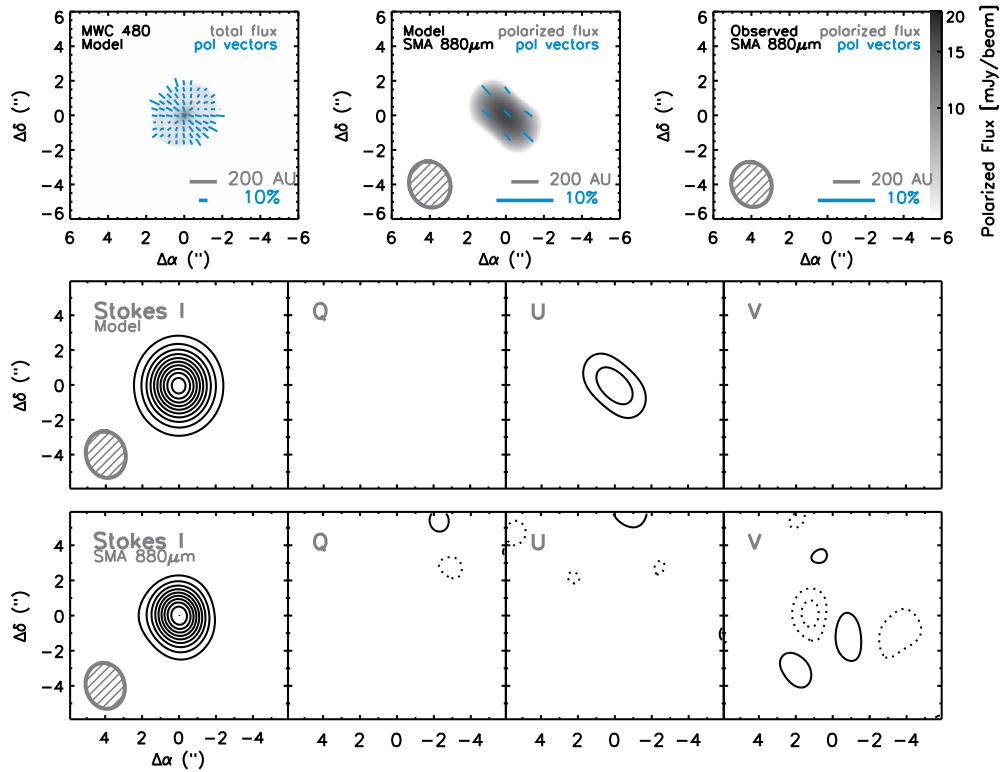}
\figcaption{ \footnotesize
Same as Figure~\ref{fig:gmaur_pol}, but for MWC~480.  The Stokes $I$ contours are multiples of 10\% of the peak flux density (0.87 \jybm), while the Stokes $Q$, $U$, and $V$ contours are in increments of 2$\sigma$, where $\sigma$ is the rms noise of 3.1 \mjybm.
\label{fig:mwc480_pol}}
\end{figure*}

\begin{figure*}[t]
\centering
\epsscale{0.9}
\plotone{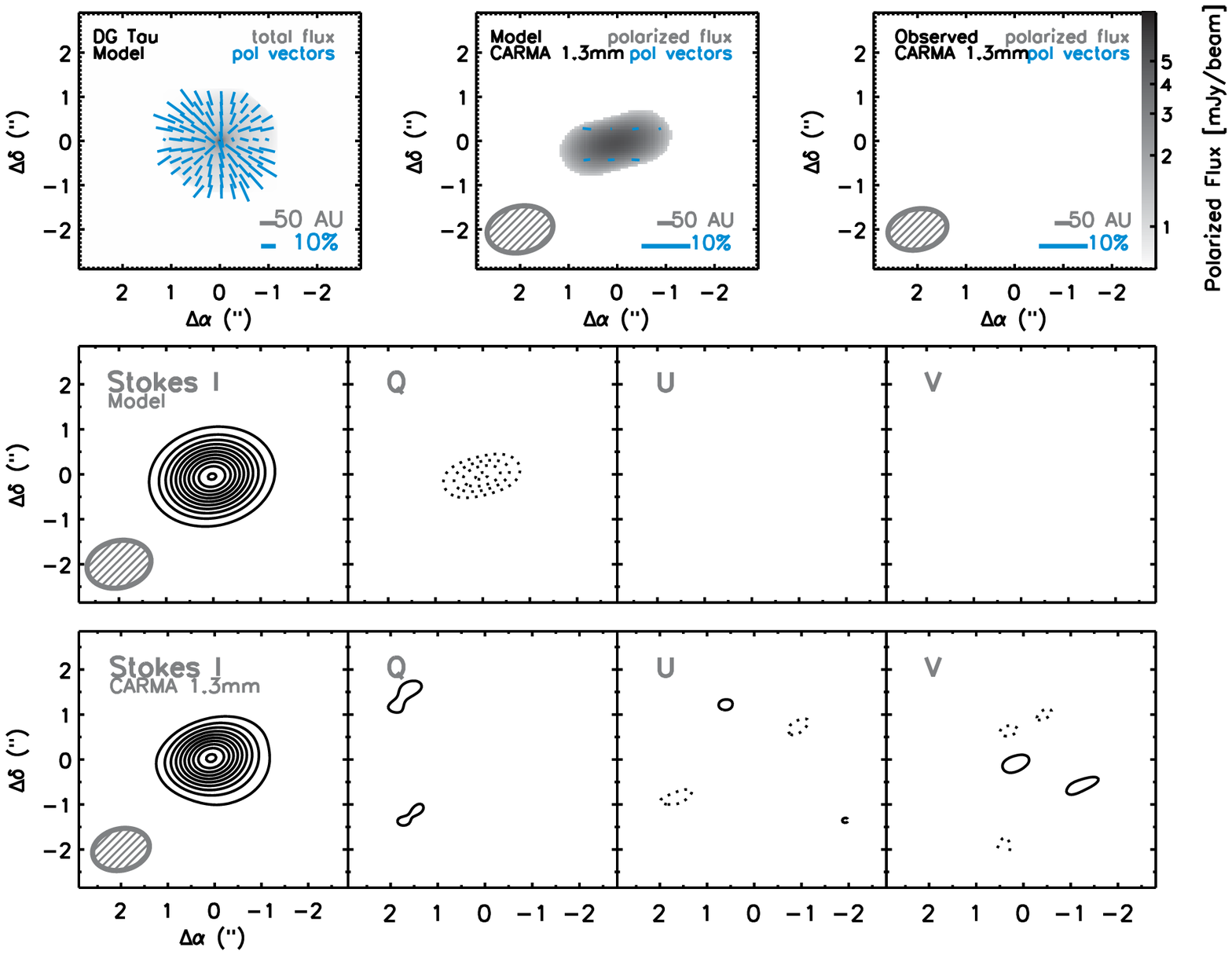}
\figcaption{ \footnotesize
Same as Figure~\ref{fig:gmaur_pol}, but for DG~Tau.  The Stokes $I$ contours are multiples of 20\% of the peak flux density (0.25 \jybm), while the Stokes $Q$, $U$, and $V$ contours are in increments of 2$\sigma$, where $\sigma$ is the rms noise of 0.66 \mjybm.
\label{fig:dgtau_pol}}
\end{figure*}

We use these models as inputs to the \citet{cho07} code, as described in \citet{hug09b}, to predict the position-dependent magnitude of polarized flux density that should be observed in Stokes $Q$ and $U$ (the models predict no circularly-polarized Stokes $V$ emission).  Figures \ref{fig:gmaur_pol}, \ref{fig:mwc480_pol}, and \ref{fig:dgtau_pol} show a comparison between the data and the polarization models produced by the \citet{cho07} code for GM~Aur, MWC~480, and DG~Tau, respectively.  Just as for the HD~163296 and TW~Hya systems presented in \citet{hug09b}, the theory substantially overpredicts the upper limit on the polarized flux density. While we would expect to detect polarized emission at the 5, 7, and 8$\sigma$ levels for MWC~480, GM~Aur, and DG~Tau, respectively, we detect no polarized continuum emission from any system.

\begin{table*}[hbt!]
\caption{Model Parameters}
\footnotesize
\begin{center}
\begin{tabular}{lcccccc}
\hline
 & \multicolumn{2}{c}{GM~Aur} & \multicolumn{2}{c}{MWC~480} & \multicolumn{2}{c}{DG~Tau} \\
Parameter & Value & Ref. & Value & Ref. & Value & Ref. \\
\hline
\hline
$T_*$ (K) & 4000 & 1 & 8460 & 2 & 3890 & 9,10 \\
$R_*$ (R$_{\sun}$) & 1.7 & 1 & 1.6 & 2 & 2.87 & 9,10 \\
$M_*$ (M$_{\sun}$) & 0.84 & 3 & 1.65 & 4 & 0.3 & 9,10 \\
$p$ & 1.1 & 5 & 1.0 & 6 & 1.0 & -- \\
$a_\mathrm{inner}$ (AU) & 20 & 5 & 0.1$^a$ & -- & 0.14 & 11 \\
$a_0$ (AU) & 300 & 5 & 275 & 6 & 89 & 10 \\
$r_\mathrm{max,i}$ ($\mu$m) & $10^3$ & 5 & $10^3$ & 6 & $10^3$ & -- \\
$i$ & 56$^\circ$ & 3 & 37$^\circ$ & 7 & 28.5$^\circ$ & 12 \\
$d$ (pc) & 140 & 8 & 140 & 8 & 140 & 8 \\
$\Sigma_0$ (g cm$^{-2}$) & 175 & -- & 275 & -- &  91 & -- \\
\hline
\end{tabular}
\tablerefs{~
{\it (a)}~Estimated dust destruction radius;
(1) \citet{bec90};
(2) \citet{ken95};
(3) \citet{dut98};
(4) \citet{sim00};
(5) \citet{hug09};
(6) \citet{hug08};
(7) \citet{ham06};
(8) \citet{eli78};
(9) \citet{muz98};
(10) \citet{ise09};
(11) \citet{pin08};
(12) \citet{ise10}
}
\end{center}
\label{tab:more_pol_model}
\end{table*}

\section{DISCUSSION}
\label{sec:discussion}

Despite observational and theoretical results implying that a 2--3\% polarization fraction may be common to circumstellar disks \citep{tam99,cho07}, recent interferometric observations of disks around pre-main sequence stars have yielded only non-detections of polarized emission \citep{hug09b,kre09}.  We have revisited both T Tauri stars with previously reported single-dish polarization measurements from the JCMT, as well as a third bright HAeBe system; these observations more than double the sample of circumstellar disks with interferometric measurements of polarization at millimeter wavelengths, and were taken across a wavelength range consistent with the \citet{tam99} results.  We report non-detections of polarized millimeter continuum emission from the disks around GM~Aur, MWC~480, and DG~Tau.  These results support the conclusions reached in \citet{hug09b}, namely that a low polarization fraction is likely common to the outer regions of bright nearby circumstellar disks.  The upper limits for GM~Aur, MWC~480, and DG~Tau are less significant than those for HD~163296 and TW~Hya because Stokes $I$ continuum emission from the new sources is fainter, and thus observations comparably sensitive in flux are less sensitive in percentage of unpolarized light.  Nevertheless, the 2--3\% polarization fractions expected from the work of \citet{cho07} and \citet{tam99} are ruled out at the 5--8$\sigma$ level by these observations, strongly suggesting that a substantially lower polarization fraction is common in circumstellar disks.  

The observations of GM~Aur and DG~Tau are particularly interesting in light of the results from \citet{tam99}.  They report polarization percentages of $3.3 \pm 1.3\%$ and $3.0 \pm 0.9\%$  for GM~Aur and DG~Tau, respectively, based on 880~$\mu$m observations with the SCUBA polarimeter on the single-dish James Clerk Maxwell Telescope (JCMT).  Although the formal significance of these detections is low (2.5$\sigma$ and 3.3$\sigma$, respectively), they note that the alignment of the polarization vector with the disk geometry is particularly suggestive of an association with circumstellar material.  Given the stringent upper limits we report here, however, the association with the disk is unlikely.  The relatively low inclination of the DG Tau system \citep[28.5$^\circ$][]{ise10} also makes such a high polarization fraction improbable.  Either the JCMT results are spurious, or the emission arises from an extended remnant envelope that is picked up by the $14\arcsec$ JCMT beam, but is spatially filtered by the SMA and CARMA so as to be undetectable.  


In light of the interferometric observations, it is unlikely that significant polarized emission is generated on large ($\gtrsim$100\,AU) scales by the disk.  This result effectively removes the observational justification for expecting an integrated 2--3\% polarization fraction in circumstellar disks.  The theoretical justification is also weakened by follow-up work to the \citet{cho07} result \citep{laz07b,hoa09}, which predicts a smaller alignment efficiency for large grains than is assumed in the original analysis \citep[see Section 4.2.2 of][]{hug09b}.  The increased number of sources in the current sample also increases the likelihood that low polarization fractions are common to T Tauri and HAeBe disks.  However, it should be noted that all the disks observed so far have been selected for their large millimeter fluxes.  They are among the very brightest and most massive circumstellar disks, and therefore may not be a representative sample.  

The interpretation presented in \citet{hug09b} still stands, namely that the low polarization fraction is most likely due to a combination of factors including rounding of dust grains, low alignment efficiency of large grains, and/or tangling of magnetic fields on large scales, possibly due to turbulence.  It also bears repeating that a low polarization fraction does not necessarily translate directly to a low magnetic field strength: in the \citet{cho07} models, there is effectively a threshold magnetic field strength above which grains will align and below which they will not.  The degree of polarization depends far more on grain shape and efficiency of alignment than on the magnetic field strength.  

The \citet{hug09} interpretation has important implications for future observations.  While several of the relevant factors (grain rounding and low alignment efficiency) predict that the polarization fraction would be uniformly low in all millimeter-wavelength observations of circumstellar disks, other factors such as magnetic field tangling offer a glimmer of hope for detection.  Turbulent motions are expected to occur on maximum size scales that are comparable to the scale height in the disk.  All the observations undertaken so far have a spatial resolution of many tens to hundreds of AU, far larger than the scale height at comparable distances from the central star.  Even if magnetic fields are ordered at some level, on these large scales any structure could be entirely washed out by integration across such a large beam.  Increasing the spatial resolution to probe regions smaller than the scale height of the disk could conceivably resolve ordered structures in the disk, thus allowing detection of polarized emission.  
In addition, higher resolution observations of the inner disk, where MRI-induced magnetic field strengths are predicted to be higher, would improve the probability of detection if, in fact, low magnetic field strengths are responsible for the lack of alignment of large grains \citep[see discussion in][]{bai11}.

\section{SUMMARY}
\label{sec:summary}

We have revisited the two T Tauri disks previously reported to exhibit significant millimeter-wavelength polarization \citep[GM Aur and DG Tau;][]{tam99} at higher spatial resolution and sensitivity, and at similar wavelengths. Furthermore, we have added a new polarimetric, millimeter-wavelength observation of the HAeBe disk around MWC 480.  We place stringent upper limits on the polarized emission from all three systems, ruling out the fiducial \citet{cho07} models at the 5--8$\sigma$ level.  These observations support previous observations of other targets \citep{hug09b}, and observations at shorter wavelengths \citep{kre09}, suggesting that the expected 2--3\% polarization fraction is not common in circumstellar disks.  Now, with a cumulative sample of three T Tauri and two HAeBe systems, it is clear that the polarization fraction is far lower, even on $\sim$100\,AU scales.  Either the previous tentative detections were spurious, or the polarization originated not from the disk but rather from a spatially extended remnant envelope or outflow.  

There are many possible explanations for the low polarization fraction, which are laid out in detail in \citet{hug09b}.  Some of the likely explanations, including rounded grains, inefficient alignment of large grains, or insufficient magnetic field strength for alignment, present a bleak prospect for future observations of polarization in circumstellar disks.  However, other possible explanations, including tangling by turbulent motions, suggest that high spatial resolution may permit the detection of ordered structures that are otherwise washed out by averaging over larger beam sizes.  Sensitive, high-resolution observations, for example with the Atacama Large Millimeter/Submillimeter Array, which is now nearing the end of its construction phase, could reveal ordered structure on scales smaller than the disk scale height.  Pushing to higher resolution would also allow investigation of magnetic field structure in the inner disk, where MRI-induced magnetic field strengths are predicted to be larger and are therefore more likely to surpass the threshold for grain alignment.  

\acknowledgments
We thank Jungyeon Cho for the use of his code to model the observations.  A.M.H. is supported by a fellowship from the Miller Institute for Basic Research in Science.  C.L.H.H. acknowledges support from an NSF Graduate Fellowship.  Support for CARMA construction was derived from the states of California, Illinois, and Maryland, the James S. McDonnell Foundation, the Gordon and Betty Moore Foundation, the Kenneth T. and Eileen L. Norris Foundation, the University of Chicago, the Associates of the California Institute of Technology, and the National Science Foundation. Ongoing CARMA development and operations are supported by the National Science Foundation under a cooperative agreement, and by the CARMA partner universities.

\bibliographystyle{apj}
\bibliography{ms}

\begin{thebibliography}{28}
\expandafter\ifx\csname natexlab\endcsname\relax\def\natexlab#1{#1}\fi

\bibitem[{{Aitken} {et~al.}(2002){Aitken}, {Efstathiou}, {McCall}, \&
  {Hough}}]{ait02}
{Aitken}, D.~K., {Efstathiou}, A., {McCall}, A., \& {Hough}, J.~H. 2002,
  \mnras, 329, 647

\bibitem[{{Andrews} {et~al.}(2009){Andrews}, {Wilner}, {Hughes}, {Qi}, \&
  {Dullemond}}]{and09a}
{Andrews}, S.~M., {Wilner}, D.~J., {Hughes}, A.~M., {Qi}, C., \& {Dullemond},
  C.~P. 2009, \apj, 700, 1502

\bibitem[{{Bai}(2011)}]{bai11}
{Bai}, X.-N. 2011, \apj, 739, 50

\bibitem[{{Beckwith} {et~al.}(1990){Beckwith}, {Sargent}, {Chini}, \&
  {Guesten}}]{bec90}
{Beckwith}, S.~V.~W., {Sargent}, A.~I., {Chini}, R.~S., \& {Guesten}, R. 1990,
  \aj, 99, 924

\bibitem[{{Cho} \& {Lazarian}(2007)}]{cho07}
{Cho}, J., \& {Lazarian}, A. 2007, \apj, 669, 1085

\bibitem[{{Dolginov} \& {Mitrofanov}(1976)}]{dol76}
{Dolginov}, A.~Z., \& {Mitrofanov}, I.~G. 1976, \apss, 43, 291

\bibitem[{{Dutrey} {et~al.}(1996){Dutrey}, {Guilloteau}, {Duvert}, {Prato},
  {Simon}, {Schuster}, \& {Menard}}]{dut96}
{Dutrey}, A., {Guilloteau}, S., {Duvert}, G., {et~al.} 1996, \aap, 309, 493

\bibitem[{{Dutrey} {et~al.}(1998){Dutrey}, {Guilloteau}, {Prato}, {Simon},
  {Duvert}, {Schuster}, \& {Menard}}]{dut98}
{Dutrey}, A., {Guilloteau}, S., {Prato}, L., {et~al.} 1998, \aap, 338, L63

\bibitem[{{Elias}(1978)}]{eli78}
{Elias}, J.~H. 1978, \apj, 224, 857

\bibitem[{{Girart} {et~al.}(2006){Girart}, {Rao}, \& {Marrone}}]{gir06}
{Girart}, J.~M., {Rao}, R., \& {Marrone}, D.~P. 2006, Science, 313, 812

\bibitem[{{Hamidouche} {et~al.}(2006){Hamidouche}, {Looney}, \&
  {Mundy}}]{ham06}
{Hamidouche}, M., {Looney}, L.~W., \& {Mundy}, L.~G. 2006, \apj, 651, 321

\bibitem[{{Hoang} \& {Lazarian}(2009)}]{hoa09}
{Hoang}, T., \& {Lazarian}, A. 2009, \apj, 697, 1316

\bibitem[{{Hughes} {et~al.}(2009{\natexlab{a}}){Hughes}, {Wilner}, {Cho},
  {Marrone}, {Lazarian}, {Andrews}, \& {Rao}}]{hug09b}
{Hughes}, A.~M., {Wilner}, D.~J., {Cho}, J., {et~al.} 2009{\natexlab{a}}, \apj,
  704, 1204

\bibitem[{{Hughes} {et~al.}(2008){Hughes}, {Wilner}, {Qi}, \&
  {Hogerheijde}}]{hug08}
{Hughes}, A.~M., {Wilner}, D.~J., {Qi}, C., \& {Hogerheijde}, M.~R. 2008, \apj,
  678, 1119

\bibitem[{{Hughes} {et~al.}(2009{\natexlab{b}}){Hughes}, {Andrews},
  {Espaillat}, {Wilner}, {Calvet}, {D'Alessio}, {Qi}, {Williams}, \&
  {Hogerheijde}}]{hug09}
{Hughes}, A.~M., {Andrews}, S.~M., {Espaillat}, C., {et~al.}
  2009{\natexlab{b}}, \apj, 698, 131

\bibitem[{{Hull} {et~al.}(2012){Hull}, {Plambeck}, {Bolatto}, {Bower},
  {Carpenter}, {Crutcher}, {Fiege}, {Franzmann}, {Hakobian}, {Heiles}, {Houde},
  {Hughes}, {Jameson}, {Kwon}, {Lamb}, {Looney}, {Matthews}, {Mundy}, {Pillai},
  {Pound}, {Stephens}, {Tobin}, {Vaillancourt}, {Volgenau}, \&
  {Wright}}]{Hull2012}
{Hull}, C.~L.~H., {Plambeck}, R.~L., {Bolatto}, A.~D., {et~al.} 2012, ArXiv
  e-prints

\bibitem[{{Isella} {et~al.}(2009){Isella}, {Carpenter}, \& {Sargent}}]{ise09}
{Isella}, A., {Carpenter}, J.~M., \& {Sargent}, A.~I. 2009, \apj, 701, 260

\bibitem[{{Isella} {et~al.}(2010){Isella}, {Carpenter}, \& {Sargent}}]{ise10}
---. 2010, \apj, 714, 1746

\bibitem[{{Kenyon} \& {Hartmann}(1995)}]{ken95}
{Kenyon}, S.~J., \& {Hartmann}, L. 1995, \apjs, 101, 117

\bibitem[{{K{\"o}nigl} \& {Salmeron}(2011)}]{kon11}
{K{\"o}nigl}, A., \& {Salmeron}, R. 2011, {The Effects of Large-Scale Magnetic
  Fields on Disk Formation and Evolution}, ed. P.~J.~V. {Garcia}, 283--352

\bibitem[{{Krejny} {et~al.}(2009){Krejny}, {Matthews}, {Novak}, {Cho}, {Li},
  {Shinnaga}, \& {Vaillancourt}}]{kre09}
{Krejny}, M., {Matthews}, T., {Novak}, G., {et~al.} 2009, ArXiv e-prints

\bibitem[{{Lazarian}(2007)}]{laz07}
{Lazarian}, A. 2007, Journal of Quantitative Spectroscopy and Radiative
  Transfer, 106, 225

\bibitem[{{Lazarian} \& {Hoang}(2007)}]{laz07b}
{Lazarian}, A., \& {Hoang}, T. 2007, \apjl, 669, L77

\bibitem[{{Marrone} \& {Rao}(2008)}]{mar08}
{Marrone}, D.~P., \& {Rao}, R. 2008, in Society of Photo-Optical
  Instrumentation Engineers (SPIE) Conference Series, Vol. 7020, Society of
  Photo-Optical Instrumentation Engineers (SPIE) Conference Series

\bibitem[{{Muzerolle} {et~al.}(1998){Muzerolle}, {Calvet}, \&
  {Hartmann}}]{muz98}
{Muzerolle}, J., {Calvet}, N., \& {Hartmann}, L. 1998, \apj, 492, 743

\bibitem[{{Pinte} {et~al.}(2008){Pinte}, {M{\'e}nard}, {Berger}, {Benisty}, \&
  {Malbet}}]{pin08}
{Pinte}, C., {M{\'e}nard}, F., {Berger}, J.~P., {Benisty}, M., \& {Malbet}, F.
  2008, \apjl, 673, L63

\bibitem[{{Simon} {et~al.}(2000){Simon}, {Dutrey}, \& {Guilloteau}}]{sim00}
{Simon}, M., {Dutrey}, A., \& {Guilloteau}, S. 2000, \apj, 545, 1034

\bibitem[{{Tamura} {et~al.}(1999){Tamura}, {Hough}, {Greaves}, {Morino},
  {Chrysostomou}, {Holland}, \& {Momose}}]{tam99}
{Tamura}, M., {Hough}, J.~H., {Greaves}, J.~S., {et~al.} 1999, \apj, 525, 832

\end{thebibliography}

\end{document}